\documentclass[12pt,amsmath,amssymb,footinbib]{revtex4-1}  %% REVTeX 4.1

\usepackage{graphicx}% Include figure files
\usepackage{dcolumn}% Align table columns on decimal point
\usepackage{bm}% bold math
\usepackage{mathrsfs}
\usepackage{natbib}

\begin{document}

\title{Energy flow and momentum under paraxial regime}% Force line breaks with \\

\author{R.~Mahillo-Isla} 
\address{\'Area de Ing. Proc. de Fab., University of Valladolid. \\Paseo del Cauce 59, 47011, Valladolid, Spain} 
\email{raumah@eii.uva.es}
\author{M.~J.~Gonz\'{a}lez-Morales}
\address{Teor\'{\i}a de la Se\~{n}al y Comunicaciones e Ingenier\'{\i}a Telem\'{a}tica. University of Valladolid.\\ Paseo de 
Bel\'{e}n 15, 47011, Valladolid, Spain} 

\date{\today}

\begin{abstract}
The paraxial model of propagation is an approximation to the model described by the d'Alembert equation. It is widely 
used to describe beam 
propagation and near-field diffraction patterns. Therefore, its use in optics and acoustics engineering is rather general. 
On the other hand, energetic balance and momentum in the electromagnetic or acoustic frameworks are well-known and 
lay in their own physical context. %, either electromagnetic or mechanical respectively. 
When dealing with paraxial solutions, these analyses are not so clear since paraxial propagation is not supported by the 
electromagnetic or mechanic theory. The present document establishes the fundamental energy and momentum analysis for 
paraxial solutions based on the classical approach by studying a Lagrangian density associated to the paraxial equation. 
Solutions of the paraxial wave equation, such as plane waves, Green's function and Gaussian beams, are studied under this 
scheme.
\end{abstract}
%Concepts: Beam optics. Traveling Waves

%\ocis{070.2580, 070.7325, 140.330, 260.1960}% REPLACE WITH CORRECT OCIS CODES FOR YOUR ARTICLE
                          % NOTE: \ocis{} IS ALIASED TO \pacs{} BUT MUST
                          % FORMAT THE TERMS CORRECTLY FOR EACH JOURNAL
\pacs{02.30.Jr, 42.60.Jf, 42.25.Bs}
%   Partial differential equations
%	Wave propagation, transmission and absorption [see also 41.20.Jb?in electromagnetism; for propagation in             %atmosphere, see 42.68.Ay; see also 52.40.Db Electromagnetic (nonlaser) radiation interactions with plasma
%and 52.38- %   r %Laser-plasma interactions?in plasma physics]
%   Beam characteristics: profile, intensity, and power; spatial pattern formation
\maketitle

\section{Introduction}

The paraxial model of propagation is widely used not only to describe beam propagation \cite{Siegman}, but also 
diffraction problems under the Fresnel approximation \cite{Grella'82,BornWolf}. Gaussian beams and their modes are its 
best known solutions and the Fresnel diffraction formula has been applied to obtain near field diffraction patterns. The 
success of paraxial propagation shows that it is a fine 
approximation since its is used use several disciplines, more or less near physics, such as material processing \cite{Ion}, 
spectroscopy \cite{RadSolPai} and medicine \cite{Welch} to cite a few. Nevertheless, up to our knowledge, the energy flow 
and momentum have never been treated within this approximation. As far as the paraxial approximation holds, the use of 
the formulae obtained from 
the full propagation framework could be a fair procedure for energy and momentum description. Nonetheless, since the 
solutions do not fulfill this scheme, the conservation of these quantities is not obtained. The classical procedure to study 
the conservation of these quantities under the d'Alembert propagation scheme is to perform a reconstruction method on the 
paraxial solution, that is, to obtain a solution of the d'Alembert equation from a paraxial solution. But there is not a unique 
method to obtain a d'Alembert solution from a paraxial one \cite{Mahillo'15,Mahillo'16}. Therefore, it is worth 
developing, if possible, the energy and momentum properties associated to the paraxial operator. The derived formulae 
should be used when dealing with solutions of the paraxial wave equation. Previous related work has focused on the 
modification of Maxwell's equations to obtain an electromagnetic description of paraxial solutions 
\cite{MukSim'85,SimSud'86} where the paraxial version of the Poynting vector is obtained. Although there is published 
material regarding energy invariants in paraxial solutions \cite{VaRuLen'07}, we have not found the treatment of 
momentum under paraxial regime neither in the scalar nor in the vectorial frameworks.

The organization of the article is as follows. First, we present the partial differential equation to be studied. Then we study a 
Lagrangian associated to this equation where the expressions of energy flow and momentum are obtained as well as their 
respective conservation equations. Based on these expressions, we also propose a method for evaluating 
the quality of a paraxial solution at every point in the space. Some fundamental solutions of the paraxial wave equation are 
considered in this article as examples: paraxial plane waves and paraxial Green's function are the classical constituents of 
any other solution in terms of plane wave spectrum and convolution integrals respectively. We also deal with the most 
popular paraxial solution, the Gaussian beam. The developed framework allows us to gain a better 
understanding of the paraxial model of propagation.

\section{Paraxial propagation}
To account not only for monochromatic waves, we use as generalized motion equation a time dependent scheme that under 
time harmonic regime turns into the classical paraxial wave equation. Thus, we deal with a wave and not with the, by far, 
more usual complex envelope. The paraxial differential equation is usually obtained by means of the application of the 
\emph{slowly varying envelope approximation} to a time harmonic solution. This leads to the parabolic equation for the 
complex envelope $\Psi$
\begin{equation}
\nabla_\perp^2\Psi+2ik\partial_z\Psi=0.
\end{equation}
In this equation the axial direction is $+z$, the time harmonic dependence is $e^{-i\omega t}$, 
the symbol $\perp$ denotes operation in the transverse plane to $z$ 
and $k=\frac{\omega}{c}$ is the wavenumber. Note that, the complex envelope is
\begin{equation}
\Psi=\tilde\psi\exp(-ikz),
\end{equation}
and thus, the equation that must be fulfilled by the phasor $\tilde\psi$ is
\begin{equation}\label{eqn:HelmP}
\nabla_\perp^2\tilde\psi+2ik\partial_z\tilde\psi+2k^2\tilde\psi=0.
\end{equation}
Since, in free space, the wavenumber is proportional to the angular frequency and the angular frequency can be identified 
as a partial derivative of the solution $\psi$ with respect to $t$; Eq. (\ref{eqn:HelmP}) comes from the generalized motion 
equation,
\begin{equation}\label{eqn:Dalpareq}
\nabla^2_\perp\psi-\frac{2}{c}\partial_{tz}\psi-\frac{2}{c^2}\partial_{tt}\psi=0,
\end{equation}
where the axial direction is $+z$ . The solutions with time harmonic dependence will be analyzed later. Although this 
equation has been obtained from a particular choice of time harmonic dependence, it does not depend on this choice. 
Conversely, the partial differential equations for the phasor and the complex envelope does depend on the particular choice 
of the time harmonic dependence. Note that there are solutions of the full wave equation that are also solutions of this 
equation, for instance, the forward propagating d'Alembert solution:
\begin{equation}\label{eqn:DFPW}
\psi=F(\mathbf{r}_\perp)g(z-ct),\;\textrm{if}\left\{\begin{array}{l} \nabla^2_\perp F=0\\g(u)\in \mathcal{C}^2\end{array}\right.
\end{equation}
This solution is not of our interest since they have been widely studied within the framework of transmission 
lines. The most important point is that such a solution represents a wave that propagates along the 
positive $z$ axis and the energy density flow at any point is well defined at any time as well as the momentum density. 
Furthermore, since Eq. (\ref{eqn:DFPW}) 
is solution of the d'Alembert equation, which is a hyperbolic partial differential equation, it is suspected that Eq. 
(\ref{eqn:Dalpareq}) should be hyperbolic as well. The matrix associated to Eq. (\ref{eqn:Dalpareq}) is
\begin{equation}
\bm{A}=\left[\begin{array}{cccc}
1&0&0&0\\
0&1&0&0\\
0&0&0&-c^{-1}\\
0&0&-c^{-1}&-2c^{-2}
\end{array}\right].
\end{equation}
The eigenvalues $\lambda_n$ of $\bm{A}$ are 
%obtained by finding the roots of the characteristic polynomial of this matrix,
%\begin{equation}
%(\lambda-1)^2\left(\lambda^2+\frac{2}{c^2}\lambda-\frac{1}{c^2}\right)=0.
%\end{equation}
%From this last equation, the eigenvalues 
\begin{subequations}
\begin{eqnarray}
\lambda_1&=&1;\\
\lambda_2&=&1;\\
\lambda_3&=&\frac{\sqrt{1+c^{-2}}-1}{c^2},\,>0;\\
\lambda_4&=&-\frac{\sqrt{1+c^{-2}}+1}{c^2}\,<0;
\end{eqnarray}
\end{subequations}
for any $c\in\mathbb{R}$. Therefore, Eq. (\ref{eqn:Dalpareq}) is a hyperbolic partial differential equation as expected. 
The main goal of this paper is to establish the general energy and momentum balance for any solution of Eq. 
(\ref{eqn:Dalpareq}) and, consequently for Eq. (\ref{eqn:HelmP}) also. For that, we study the Lagrangian associated to Eq. 
(\ref{eqn:Dalpareq}).

\section{Analysis of the Lagrangian density of the paraxial equation}
The Lagrangian density associated to Eq. (\ref{eqn:Dalpareq}) can be written by simple inspection of the differential 
equation,
\begin{equation}\label{eqn:Lagrangian}
\mathscr{L}=\frac{1}{c}\partial_t\psi\left(\frac{1}{c}\partial_t\psi+\partial_z\psi\right)-
\frac{1}{2}\left[(\partial_x\psi)^2+(\partial_y\psi)^2\right]
\end{equation}
In fact, we can recover Eq. (\ref{eqn:Dalpareq}) by applying Euler-Lagrange equation
\begin{equation}\label{eqn:ELeq}
\nabla_{4}\cdot\left(\sum_{\mu=x,y,z,t}\frac{\partial\mathscr{L}}
{\partial(\partial_\mu\psi)}\mathbf{e}_\mu\right)=\frac{\partial\mathscr{L}}{\partial\psi},
\end{equation}
where $\nabla_4=\mathbf{e}_t\partial_t+\nabla$ in terms of the generalized coordinates. Let us calculate the terms:
\begin{equation}
\sum_{\mu=x,y,z,t}\frac{\partial\mathscr{L}}{\partial(\partial_\mu\psi)}\mathbf{e}_\mu=
\frac{1}{c}\left(2\frac{\partial_t\psi}{c}+\partial_z\psi\right)\mathbf{e}_t-\nabla_\perp\psi+
\frac{\partial_t\psi}{c}\mathbf{e}_z,
\end{equation}
and the left hand side of Eq. (\ref{eqn:ELeq}) produces
\begin{equation}
\frac{2}{c^2}\partial_{tt}\psi+\frac{1}{c}\partial_{zt}\psi-\nabla_\perp^2\psi+\frac{1}
{c}\partial_{tz}\psi=\frac{\partial\mathscr{L}}{\partial\psi}.
\end{equation}
By requiring the necessary conditions of regularity on $\psi$, $\partial_{zt}\psi=\partial_{tz}\psi$, 
\begin{equation}
\nabla_\perp^2\psi-\frac{2}{c}\partial_{tz}\psi-\frac{2}{c^2}\partial_{tt}\psi=
-\frac{\partial\mathscr{L}}{\partial\psi}.
\end{equation}
From Eq. (\ref{eqn:Lagrangian}), it follows that
\begin{equation}
\frac{\partial\mathscr{L}}{\partial{\psi}}=0,
\end{equation}
and the expression in Eq. (\ref{eqn:Dalpareq}) is recovered. Since the Lagrangian does not depend on the coordinates 
explicitly, the hypothesis of Noether's theorem are fulfilled. Thus, there are four primary conserved quantities. One of them 
is the energy and the other three ones are momentum. The conservation of the energy is found in the ``time'' row (usually 
the first row) of the stress-energy tensor. This row is the four-component vector
\begin{equation}
\mathbf{T}_t=\partial_t\psi\left(\sum_{\mu=x,y,z,t}\frac{\partial\mathscr{L}}
{\partial(\partial_\mu\psi)}\mathbf{e}_{\mu}\right)-\mathscr{L}\mathbf{e}_t.
\end{equation}
The energy density $U$ is identified as the t-component of $\mathbf{T}_t$,
\begin{equation}\label{eqn:Uparaxial}
U=\frac{(\partial_t\psi)^2}{c^2}+\frac{1}{2}\left[(\partial_x\psi)^2+(\partial_y\psi)^2\right].
\end{equation}
The rest of the components of $\mathbf{T}_t$ forms the energy density flow $\mathbf{S}$,
\begin{equation}
\mathbf{T}_t=\mathbf{S}+U\mathbf{e}_t.
\end{equation}
Thus, the energy density flow is calculated as
\begin{equation}\label{eqn:Sparaxial}
\mathbf{S}=-\partial_t\psi\left(\nabla_\perp\psi-\frac{\partial_t\psi}
{c}\mathbf{e}_z\right),
\end{equation}
Upon Noether's theorem
\begin{equation}\label{eqn:poynpar}
\nabla\cdot\mathbf{S}+\partial_tU=0.
\end{equation}
The rest of the stress-energy tensor allows us to define the conservation of momentum. The momentum density is 
\begin{equation}\label{eqn:momentumdef}
\mathbf{p}=\left[\frac{2}{c^2}(\partial_t\psi)+\frac{1}{c}(\partial_z\psi)\right]\nabla\psi.
\end{equation}
The momentum density flow is a three tensor
\begin{equation}
\bm{M}=[\mathbf{M}_x,\mathbf{M}_y,\mathbf{M}_z]
\end{equation}
with columns
\begin{eqnarray}
\mathbf{M}_x&=&-(\partial_x\psi)\left[\nabla_\perp\psi-\frac{1}{c}
(\partial_t\psi)\mathbf{e}_z\right]-\mathscr{L}\mathbf{e}_x,\\
\mathbf{M}_y&=&-(\partial_y\psi)\left[\nabla_\perp\psi-\frac{1}{c}
(\partial_t\psi)\mathbf{e}_z\right]-\mathscr{L}\mathbf{e}_y,\\
\mathbf{M}_z&=&-(\partial_z\psi)\nabla_\perp\psi-\frac{1}{c^2}(\partial_t\psi)\mathbf{e}_z-\mathscr{L}\mathbf{e}_z.
\end{eqnarray}
Noether's theorem states that
\begin{equation}
\partial_t\mathbf{p}=-\nabla\bm{M},
\end{equation}
where $\nabla$ is a row vector in this last equation.

%Angular momentum density is derived from the momentum density by 
%its definition.
%\begin{equation}
%\mathbf{l}=\mathbf{r}\times\mathbf{p}
%\end{equation}
%Note that contrary to the case of the full d'Alembert equation
%\begin{equation}
%\mathbf{l}\neq\frac{1}{c^2}\mathbf{S}\times\mathbf{r}.
%\end{equation}
%One of the most useful calculation is the $z$ component of the angular momentum. In this case, the angular momentum 
%density with regard to the $z$ axis in cylindrical coordinates is,
%\begin{equation}
%l_z=(\partial_\phi\psi)\left[\frac{2}{c^2}(\partial_t\psi)+\frac{1}{c}(\partial_z\psi)\right].
%\end{equation}
%This quantity is conserved by means of the angular momentum density flow $\mathbf{M}_\phi$, whose expression is
%\begin{equation}
%\mathbf{M}_\phi=-(\partial_\rho\psi)(\partial_\phi\psi)\mathbf{e}_\rho+\left[\frac{1}
%{2}\left(\partial_\rho\psi-\frac{\partial_\phi\psi}{\rho^2}\right)-\frac{\partial_t\psi}{c}\left(\frac{\partial_t\psi}
%{c}+\partial_z\psi\right)\right]\mathbf{e}_\phi+\frac{\partial_t\psi}{c}\partial_\phi\psi\mathbf{e}_z.
%\end{equation} 
%The conservation equation of the angular momentum around the $z$ axis based on the Noether's theorem is
%\begin{equation}
%\frac{1}{\rho}\partial_\rho(\rho\mathbf{M}_\phi\cdot\mathbf{e}_\rho)+
%\partial_\phi(\mathbf{M}_\phi\cdot\mathbf{e}_\phi)
%+\partial_z(\mathbf{M}_\phi\cdot\mathbf{e}_z)+\partial_tl_z=0
%\end{equation}
Momentum analysis can be taken further than the coverage given in this document. Other relevant analysis, such as angular 
momentum, derive directly from ours since angular momentum density is derived from the momentum density by its 
definition.
\begin{equation}
\mathbf{l}=\mathbf{r}\times\mathbf{p}
\end{equation}
%Note that contrary to the case of the full d'Alembert equation
%\begin{equation}
%\mathbf{l}\neq\frac{1}{c^2}\mathbf{S}\times\mathbf{r}.
%\end{equation}
One of the most useful calculation could be the $z$ component of the angular momentum \cite{Allen}. In this case, the 
angular momentum density with regard to the $z$ axis in cylindrical coordinates is,
\begin{equation}
l_z=(\partial_\phi\psi)\left[\frac{2}{c^2}(\partial_t\psi)+\frac{1}{c}(\partial_z\psi)\right].
\end{equation}
This quantity is conserved by means of the angular momentum density flow $\mathbf{M}_\phi$, whose expression is
\begin{equation}
\mathbf{M}_\phi=-(\partial_\rho\psi)(\partial_\phi\psi)\mathbf{e}_\rho+\left[\frac{1}
{2}\left(\partial_\rho\psi-\frac{\partial_\phi\psi}{\rho^2}\right)-\frac{\partial_t\psi}{c}\left(\frac{\partial_t\psi}
{c}+\partial_z\psi\right)\right]\mathbf{e}_\phi+\frac{\partial_t\psi}{c}\partial_\phi\psi\mathbf{e}_z.
\end{equation} 
The conservation equation of the angular momentum around the $z$ axis is
\begin{equation}
\frac{1}{\rho}\partial_\rho(\rho\mathbf{M}_\phi\cdot\mathbf{e}_\rho)+
\partial_\phi(\mathbf{M}_\phi\cdot\mathbf{e}_\phi)
+\partial_z(\mathbf{M}_\phi\cdot\mathbf{e}_z)+\partial_tl_z=0
\end{equation}
We give the previous equation to help further studies with regard to orbital angular momentum of light beams.\\

By comparing Eqs. (\ref{eqn:momentumdef}) and (\ref{eqn:Sparaxial}) it can be seen that $\mathbf{S}$ and 
$\mathbf{p}$ are not antiparallel, in general, within the paraxial framework. Note that this is a key difference with regard 
to d'Alembert equation, as shown in Eqs. (\ref{eqn:SFW}) and (\ref{eqn:pfw}) where energy density flow and 
momentum have a simple relation of proportionality between them. A solution of the paraxial wave equation can be 
considered as a tool for describing propagation if it behaves approximately as a solution of the d'Alembert wave equation. 
Therefore, any parameter related to $\mathbf{S}$ and/or $\mathbf{p}$ can be seen as a measure of the quality of a 
paraxial solution. For instance,
\begin{equation}\label{eqn:defD}
\mathbf{D}=c^2\mathbf{p}+\mathbf{S}
\end{equation}
will measure where a solution of the paraxial wave equation can be a good approximation to wave propagation. Note that 
the last quantity is identically zero for any solution of the d'Alembert equation. Thus, the paraxial quality of a solution 
can be obtained by comparing $\mathbf{D}$ with $\mathbf{S}$; for instance
\begin{equation}\label{eqn:defQ}
Q=\frac{\|\mathbf{D}\|}{\|\mathbf{S}\|}.
\end{equation}

\section{Time harmonic solutions}
We let $\psi$ have a time harmonic dependence. Hence,
\begin{equation}\label{eqn:defprth}
\psi(\mathbf{r},t)=\tilde{\psi}(\mathbf{r})e^{-i\omega t},
\end{equation}
where $\tilde\psi(\mathbf{r})$ is a solution of (\ref{eqn:HelmP}). To recover the practical solution of (\ref{eqn:Dalpareq}) 
one has to extract, for example, the real part of (\ref{eqn:defprth}),
\begin{equation}
\psi=\Re\left\{\tilde\psi e^{-i\omega t}\right\}=\frac{1}{2}\left(\tilde\psi\,e^{-i\omega t}+\tilde\psi^*e^{i\omega 
t}\right),
\end{equation} 
where the asterisk superscript denotes the complex conjugate. This form of expressing the paraxial solution leads to a 
straightforward calculation of time average energy and flow densities. As usual, time average value of a given time 
harmonic function $A(t)$ is obtained by doing the integral
\begin{equation}
\langle A\rangle=\frac{1}{T}\int_{<T>}\mathrm{d}t\,A(t),
\end{equation}
where $T=\frac{2\pi}{\omega}$. Let us start from energy density flow calculation. First we find that
\begin{equation}
\partial_t\psi=\frac{\omega}{2i}\left(\tilde\psi\,e^{-i\omega t}-\tilde\psi^*e^{i\omega t}\right).
\end{equation}
The expressions of the partial derivatives with regard to the spatial coordinates are also needed:
\begin{equation}
\partial_\mu\psi=\frac{1}{2}\left(\partial_\mu\tilde\psi\,e^{-i\omega t}+\partial_\mu\tilde\psi^*e^{i\omega t}\right), 
\quad \mu=x,y,z.
\end{equation}
Thus, energy density flow for time harmonic solutions is
\begin{equation}
\mathbf{S}=\frac{\omega}{2}\left[
\Im\left\{\tilde\psi^*\left(\nabla_\perp\tilde\psi+ik\tilde\psi\mathbf{e}_z\right)\right\}-
\Im\left\{\tilde\psi\left(\nabla_\perp\tilde\psi+ik\tilde\psi\mathbf{e}_z\right)e^{-i2\omega t}\right\}\right]
\end{equation}
The last expression is adequate for calculating time average density flow
\begin{equation}\label{eqn:SavP}
\langle\mathbf{S}\rangle=\frac{\omega}{2}
\Im\left\{\tilde\psi^*\left(\nabla_\perp\tilde\psi+ik\tilde\psi\mathbf{e}_z\right)\right\}.
\end{equation}
The same procedure is performed with respect to the energy density $U$. The quantities involved in the calculation have 
been already obtained, and therefore
\begin{equation}
U=\frac{k^2}{2}|\tilde\psi|^2+\frac{1}{4}\left\|\nabla_\perp\tilde\psi\right\|^2
+\frac{1}{2}\Re\left\{\left\{\frac{1}{2}\left[(\partial_x\tilde\psi)^2+(\partial_y\tilde\psi)^2\right]-k^2\tilde\psi^2\right\} 
e^{-i2\omega t}\right\}.
\end{equation}
Time average energy density for time harmonic solutions is obtained from this last expression as,
\begin{equation}
\langle U\rangle=\frac{1}{2}\left(k^2|\tilde\psi|^2+\frac{1}{2}\left\|\nabla_\perp\tilde\psi\right\|^2\right).
\end{equation}
The formalism followed gives rise to the definition of a complex flow density for $\tilde\psi$ 
\begin{equation}\label{eqn:defSc}
\mathbf{S}_c=-\frac{i\omega}{2}\tilde\psi^*(\nabla_\perp\tilde\psi+ik\tilde\psi\,\mathbf{e}_z.).
\end{equation}
%where $P$ can be called generalized pressure phasor
%\begin{equation}\label{eqn:defPres}
%P=i\omega\tilde\psi,
%\end{equation}
%and $\mathbf{V}$ is, by the same token, the generalized velocity phasor
%\begin{equation}\label{eqn:defVel}
%\mathbf{V}=\nabla_\perp\tilde\psi+ik\tilde\psi\,\mathbf{e}_z.
%\end{equation}
It follows from its definition that
\begin{equation}
\langle\mathbf{S}\rangle=\Re\{\mathbf{S}_c\}.
\end{equation}
The imaginary part of $\mathbf{S}_c$ will lead to the reactive energy balance associated to the solution $\psi$. 
%Thus, for a given solution to Eq. (\ref{eqn:HelmP}), generalized pressure and velocity can be constructed in order to 
%obtain $\mathbf{S}_c$ and, consequently the energy balance of such solution.

Momentum density can also be calculated for time harmonic solutions explicitly. The procedure is quite similar to the one 
carried out to obtain $\mathbf{S}$ and yields
\begin{equation}
\mathbf{p}=\frac{\omega}{c^2}\left[\Im\left\{\tilde\psi\nabla\tilde\psi^*+\tilde\psi\nabla\tilde\psi e^{-i2\omega 
t}\right\}\right]+ \frac{1}{2c}\left[\Re\left\{\partial_z\tilde\psi\nabla\tilde\psi^*+\partial_z\tilde\psi\nabla\tilde\psi
e^{-i2\omega t}\right\}\right].
\end{equation}
From this last equation, the average momentum density is
\begin{equation}\label{eqn:pavaramP}
\langle\mathbf{p}\rangle=\frac{\omega}{c^2}\Im\left\{\tilde\psi\nabla\tilde\psi^*\right\}
+\frac{1}{2c}\Re\left\{\partial_z\tilde\psi\nabla\tilde\psi^*\right\}.
\end{equation}

Since energy density flow and momentum density do not follow a simple proportionality relation in the paraxial 
approximation scheme, one issue about using raytracing techniques may arise within this framework. In 
the full wave propagation scheme, it can be shown how, for time harmonic regime, the phase of a solution of the Helmholtz 
equation contains some information about the vector behavior of average momentum density and energy density flow 
(see Appendix \ref{susec:con}). From Eqs. (\ref{eqn:SavP}) and (\ref{eqn:pavaramP}), it can be seen how average energy 
flow and momentum densities depend on the phase of $\tilde\psi$. The decomposition of the phasor in terms of real 
amplitude $A(\mathbf{r})$ and real phase $\phi(\mathbf{r})$ (the extraction of the constant $\phi_0$ allows us to define 
$A$ and $\phi$ as real functions),
\begin{equation}
\tilde\psi=A(\mathbf{r})\exp\{i[k\phi(\mathbf{r})+\phi_0]\},
\end{equation}
in the aforementioned equations, gives the dependence of the average energy density flow and momentum density in terms 
of amplitudes and phase of $\tilde\psi$. For the average energy density flow, we obtain
\begin{equation}
\langle\mathbf{S}\rangle=\frac{\omega^2}{2c}
\left[A(\mathbf{r})\nabla_\perp\phi(\mathbf{r})+A^2(\mathbf{r})\mathbf{e}_z\right].
\end{equation}
The average momentum density in terms of amplitude and phase is
\begin{equation}\label{eqn:pAth}
\langle\mathbf{p}\rangle=\frac{1}{2c}
\left\{k^2[\partial_z\phi(\mathbf{r})-2A(\mathbf{r})]\nabla\phi(\mathbf{r})+\partial_z A(\mathbf{r})\nabla 
A(\mathbf{r})\right\}
\end{equation}
These last equations show that neither $\langle\mathbf{S}\rangle$ nor $\langle\mathbf{p}\rangle$ are proportional, in 
general, to $\nabla\phi$. Since we usually have the solution $\tilde\psi$ in this framework, characteristics of 
$\langle\mathbf{S}\rangle$ and $\langle\mathbf{p}\rangle$ cannot be inferred by studying $\nabla\phi$. The energy 
density flow depends on $\nabla_\perp\phi$ instead of $\nabla\phi$. The momentum density does depend on $\nabla\phi$ 
but there is another term that contributes with $\nabla A$. Thus, if the following condition is met,
\begin{equation}\label{eqn:cond1}
k^2[\partial_z\phi(\mathbf{r})-2A(\mathbf{r})]\gg\partial_z A(\mathbf{r}),
\end{equation}
the study of $\nabla\phi$ will help us to see features of the momentum density approximately. The condition expressed in 
this last equation requires the amplitude $A(\mathbf{r})$ to be a weak function of the coordinates and the high 
frequency condition $k\gg1$.

\section{Examples}
\subsection{Generic traveling wave towards $+z$}\label{subsec:zt}
Let us compute energy density flow with Eq. (\ref{eqn:Sparaxial}) of the 
solution in Eq. (\ref{eqn:DFPW});
%\begin{equation}
%U=F^2(\mathbf{r}_\perp)g'(z-ct)+\frac{1}{2}
%\left[\left(\partial_xF(\mathbf{r}_\perp)\right)^2+\left(\partial_yF(\mathbf{r}_\perp)\right)^2\right] g^2(z-ct).
%\end{equation}
%The energy density flow is obtained as
\begin{equation}
\mathbf{S}=c\,F(\mathbf{r}_\perp)g'(z-ct)\left[g(z-ct)\nabla_\perp F(\mathbf{r}_\perp)+F(\mathbf{r}_\perp)
g'(z-ct)\mathbf{e}_z\right].
\end{equation}
Also, the momentum density can be obtained
\begin{equation}
\mathbf{p}=\frac{-1}{c}F(\mathbf{r}_\perp)g'(z-ct)\left[g(z-ct)\nabla_\perp F(\mathbf{r}_\perp)+F(\mathbf{r}_\perp)
g'(z-ct)\mathbf{e}_z\right].
\end{equation}
These expressions are also obtained if Eqs. (\ref{eqn:SFW}) and (\ref{eqn:pfw}) are used instead. Note 
that, since the solution in (\ref{eqn:DFPW}) is also a solution to Eq. (\ref{eqn:Daleq}) the procedure explained can be done 
as well. It follows that the energy and momentum balance obtained in both cases is exactly the same, as it should be.  
%Thus, the quality $Q$ of a paraxial solution as
%\begin{equation}
%Q=\frac{-\mathbf{S}\cdot\mathbf{p}}{||\mathbf{S}||\,||\mathbf{p}||}
%\end{equation}

\subsection{Paraxial plane waves}
Under the time dependence $e^{-i\omega t}$, the phasor associated to a plane wave is
\begin{equation}
\tilde\psi=\exp(i\mathbf{k}\cdot\mathbf{r}),
\end{equation}
where
\begin{equation}
\mathbf{k}=k\left[q\left(\cos\beta\,\mathbf{e}_x+\sin\beta\,\mathbf{e}_y\right)+
\left(1-\frac{q^2}{2}\right)\mathbf{e}_z\right].
\end{equation}
Phase fronts are planes normal to the direction of $\mathbf{k}$. The zenith angle of $\mathbf{k}$ (angle with the $+z$ 
axis) is
\begin{equation}
\alpha=\arccos\left(\frac{1-\frac{q^2}{2}}{\sqrt{1+\frac{q^4}{4}}}\right),
\end{equation}
and the azimuthal angle is $\beta$. 
%The generalized pressure for this solution is directly obtained and the generalized velocity 
%\begin{equation}
%\mathbf{V}=ik\tilde\psi[q(\cos\beta\,\mathbf{e}_x+\sin\beta\,\mathbf{e}_y)+\mathbf{e}_z].
%\end{equation}
The complex energy density flow is a constant vector for such solution
\begin{equation}
%\mathbf{S}_c=\frac{k\omega}{2}[q(\cos\beta\,\mathbf{e}_x+\sin\beta\,\mathbf{e}_y)+\mathbf{e}_z]
\mathbf{S}_c=\frac{k^2c}{2}\left[\mathbf{n}+\frac{q^2}{2}\mathbf{e}_z\right],
\end{equation}
where 
\begin{equation}
\mathbf{n}=\frac{\mathbf{k}}{\|\mathbf{k}\|},
\end{equation}
 as happens with propagative usual plane waves. For real parameters $q$ and $\beta$, the vector is a 
real one and there is not reactive energy density. The average momentum density can also be computed:
\begin{equation}
\langle\mathbf{p}\rangle=\frac{-k^2}{2c}\left(1+\frac{q^2}{2}\right)\mathbf{n}.
\end{equation}
For this kind of solutions with constant amplitude, the average momentum density is parallel to $\mathbf{k}$ as  
predicted by Eq. (\ref{eqn:pAth}). Conversely, the zenith angle of $\mathbf{S}_c$ is 
\begin{equation}
\alpha_\mathbf{S}=\arccos\left(\frac{1}{\sqrt{1+q^2}}\right)
\end{equation}
which, unlike usual homogeneous Helmholtz plane waves, is different from $\alpha$. The angle $\alpha_\mathbf{S}$ is in 
$[0,\frac{\pi}{2})$ as $q$ is in $[0,+\infty)$ (this set of $q$ leads to a basis in terms of the Fourier transform for the 
paraxial equation). In other words, $\mathbf{S}_c$ cannot have components towards $-z$. The norms of both 
$\langle\mathbf{p}\rangle$ and $\mathbf{S}_c$ are functions of the direction which is also a difference with respect to 
Helmholtz plane waves. This fact can be explained because the medium gets denser as the zenith angle of $\mathbf{k}$ 
increases, 
\begin{equation}
\|\mathbf{k}\|=k\sqrt{1+\frac{q^4}{4}}.
\end{equation}
As a consequence, the phase velocity
\begin{equation}\label{eqn:natconvf}
v_p\leq c,
\end{equation}
for paraxial plane waves. Besides, except for $q=0$ constant phase curves are not orthogonal to energy density flow as was 
previously predicted. Momentum density can be written in terms of energy density flow as,
\begin{equation}
\langle\mathbf{p}\rangle=\frac{1}{c^{2}}\left(\mathbf{D}-\mathbf{S}_c\right)
\end{equation}
where $\mathbf{D}$ is obtained as
\begin{equation}
\mathbf{D}=\frac{k^2c}{2}\frac{q^2}{2}\left(\mathbf{n}-\mathbf{e}_z\right).
\end{equation}
Note that as $q$ increases, the vector $\mathbf{D}$ gets larger, 
showing that the idea of measuring the quality of the paraxial solution based on this vector is a fair 
procedure. The case $q=0$ is a special case of the example in Subsection \ref{subsec:zt}.

\subsection{Green's function of the time harmonic paraxial wave equation}
Green's function associated to the operator in Eq. (\ref{eqn:HelmP}) is
\begin{equation}
\tilde\psi=\frac{1}{4\pi z}\exp\left[ik\left(z+\frac{\rho^2}{2z}\right)\right].
\end{equation}
The complex energy density flow can be obtained by using Eq. (\ref{eqn:defSc}),
% (\ref{eqn:defPres}) and 
%(\ref{eqn:defVel}). The expression of the generalized pressure is 
%\begin{equation}
%P=i\omega\tilde\psi=\frac{i\omega}{4\pi z}\exp\left[ik\left(z+\frac{\rho^2}{2z}\right)\right].
%\end{equation}
%The generalized velocity field can be written as
%\begin{equation}
%\mathbf{V}=\frac{ik}{4\pi z^2}\exp\left[ik\left(z+\frac{\rho^2}{2z}\right)\right]\mathbf{r}.
%\end{equation}
%The complex energy density flow is 
\begin{subequations}
\begin{eqnarray}\label{eqn:seSGC}
\mathbf{S}_c&=&\frac{k^2c}{32\pi^2}\frac{1}{z^2}\left(\frac{\rho}{z}\mathbf{e}_\rho+\mathbf{e}_z\right),
\quad(\textrm{cyl.});\\ \label{eqn:seSGS}
\mathbf{S}_c&=&\frac{k^2c}{32\pi^2}\frac{1}{r^2\cos^3\theta}\mathbf{e}_r,\quad (\textrm{sph.}).
\end{eqnarray}
\end{subequations}
There is not reactive energy in this solution since the complex energy density flow is a real quantity. As shown in Eq. 
(\ref{eqn:seSGS}), field lines of $\mathbf{S}_c$ are radial. In the semispaces of $z<0$, the energy goes towards 
the origin, where it concentrates and then, the energy is radiated towards the $z>0$ semispace. Formally, this statement 
can be shown by means of the divergence. It is easy to see that, in general,
\begin{equation}\label{eqn:div0}
\nabla\cdot\mathbf{S}_c=0.
\end{equation}
From the viewpoint of the energy, this last equation shows that there is not energy storage, only energy flow. 
The origin in such solutions must be treated by means of the definition of the divergence. Let $B_R(0)$ be a ball 
with center at the origin and radius $R$. Then
\begin{equation}
\left.\nabla\cdot\mathbf{S}_c\right|_{\mathrm{origin}}=\lim_{R\to0}\frac{3}{4\pi}\frac{1}{R^3}\iint_{\partial B_R(0)}
\mathrm{d}a\,\mathbf{S}_c\cdot\mathbf{e}_r.
\end{equation}
Using spherical coordinates, 
\begin{equation}
\left.\nabla\cdot\mathbf{S}_c\right|_{\mathrm{origin}}=\frac{3k^2c}{128\pi^3}
\lim_{R\to0}\frac{1}{R^3}\int_{0}^{\pi}R\mathrm{d}\theta\int_{0}^{2\pi}R\sin\theta\mathrm{d}\varphi\,
\frac{1}{R^2\cos^3\theta}.
\end{equation}
By operating in this last equation, it is found that,
\begin{equation}
\left.\nabla\cdot\mathbf{S}_c\right|_{\mathrm{origin}}=\frac{3k^2c}{64\pi^2}\lim_{R\to0}\frac{1}{R^3}
\int_{0}^{\pi}\mathrm{d}\theta\,\frac{\sin\theta}{\cos^3\theta}.
\end{equation}
The integral in this last equation is equal to zero and thus,
\begin{equation}
\left.\nabla\cdot\mathbf{S}_c\right|_{\mathrm{origin}}=0,
\end{equation}
showing that Eq. (\ref{eqn:div0}) holds for every point in the space. This solution is a very special one because, from the 
viewpoint of the energy, it lacks any source even though it was generated by an impulse function. Therefore, by means of 
the convolution integral, any radiation problem in time harmonic regime is automatically transformed into a propagation 
problem without sources.
%In other words, the paraxial propagation scheme is not suitable for radiation problems in general.
It follows, that this feature is also naturally inherited by solutions of general time dependence. 

The average momentum density of Green's function can be obtained as
\begin{equation}\label{eqn:avmdg}
\langle\mathbf{p}\rangle=\frac{-k^2}{32c\pi^2}\frac{1}{z^2}\left(1+\frac{\rho^2}{2z^2}\right)
\left[\frac{\rho}{z}\mathbf{e}_\rho+\left(1-\frac{\rho^2}{2z^2}-\frac{k^{-2}}{z^2
+\frac{\rho^2}{2}}\right)\mathbf{e}_z\right]. 
\end{equation}
Note that, it is easy to write $\langle\mathbf{p}\rangle$ in terms of $\mathbf{S}_c$ plus another term that accounts for 
the difference between them,
\begin{equation}\label{eqn:pvsS}
\langle\mathbf{p}\rangle=-\frac{1}{c^2}\mathbf{S}_c+\frac{k^2}{32c\pi^2}\frac{1}{z^2}
\left[\frac{-\rho^3}{2z^3}\mathbf{e}_\rho+\left(\frac{\rho^4}{4z^4}+\frac{k^{-2}}{z^2+\frac{\rho^2}{2}}
+\frac{k^{-2}\rho^2}{2z^4+z^2\rho^2}\right)\mathbf{e}_z\right].
\end{equation}
Therefore, the proposed vectorial quantity defined in Eq. (\ref{eqn:defD}) is
\begin{equation}
\mathbf{D}=\frac{k^2c}{32\pi^2}\frac{1}{z^2}
\left[\frac{-\rho^3}{2z^3}\mathbf{e}_\rho+\left(\frac{\rho^4}{4z^4}+\frac{k^{-2}}{z^2+\frac{\rho^2}{2}}
+\frac{k^{-2}\rho^2}{2z^4+z^2\rho^2}\right)\mathbf{e}_z\right].
\end{equation}
This last equation shows that momentum density and energy density flow are not antiparallel in general. In high frequency 
conditions ($k\gg 1$), $\mathbf{D}$ can be neglected with regard to $\mathbf{S}_c$ in the regions 
where $|z|\gg\rho$, that is, points near the axial direction. This is a well known result since the paraxial Green's function 
can be obtained by the appropriate approximations of the distance in the Helmholtz Green's function. Again, this result 
shows that the comparison between momentum density and energy density flow gives a fair procedure to obtain a 
measure of the quality of a paraxial solution. Last equations are rather cumbersome to establish further analytical 
differences between momentum density and energy density flow. Nonetheless, since we know that energy density flow of 
Green's function is radial, these differences can be seen graphically.
\begin{figure}[!htb]
\begin{center}
\includegraphics[width=8.6cm]{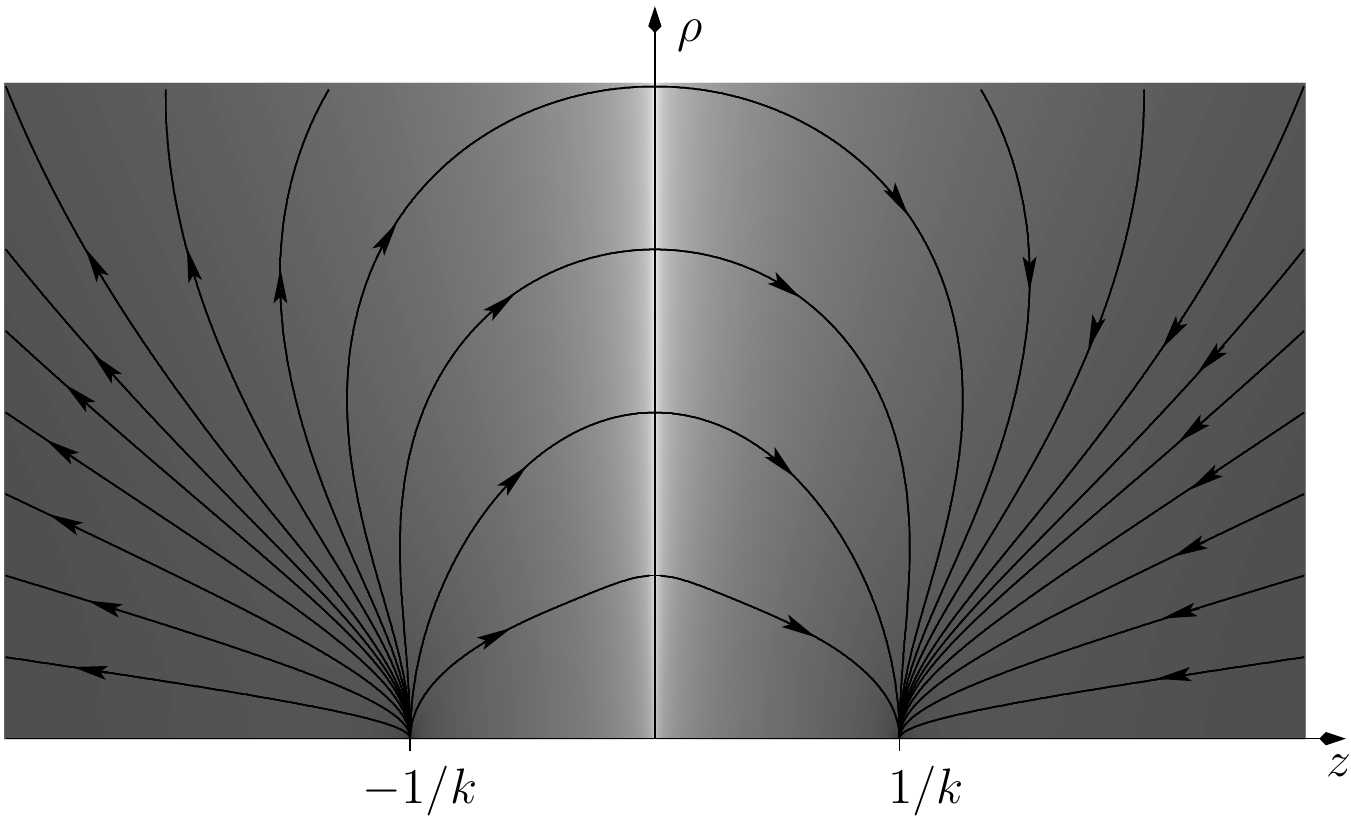}
\caption{Momentum density of paraxial Green's function: Field lines and 
intensity. The level of gray indicates the norm of the momentum density at that point, being black the 
minimum values (at $z=\pm k^{-1}$ in the z axis) and white the maximum ones (concentrated in the $z=0$ plane). The 
intensity is in logarithmic scale.}\label{fig:AMPGreen}
\end{center}
\end{figure}
In Fig. \ref{fig:AMPGreen} a sketch of momentum density of paraxial Green's function is shown. All the field lines start at 
the point $(\rho=0,z=-k^{-1})$ and end at $(\rho=0,z=k^{-1})$ and cross the $z=0$ plane in its normal direction. At 
this plane, momentum density and energy density flow are ill-defined just like the phasor. No other points, except those 
fulfilling $|z|\gg\rho$ and $|z|>k^{-1}$ can be seen as a region where the paraxial Green's function is a fair tool 
for representing wave propagation. Note that the origin is excluded from this region since near this region 
$\mathbf{S}_c$ and $\langle\mathbf{p}\rangle$ are parallel and not antiparallel.  

\begin{figure}[!htbp]
\begin{center}
\includegraphics[width=8.6cm]{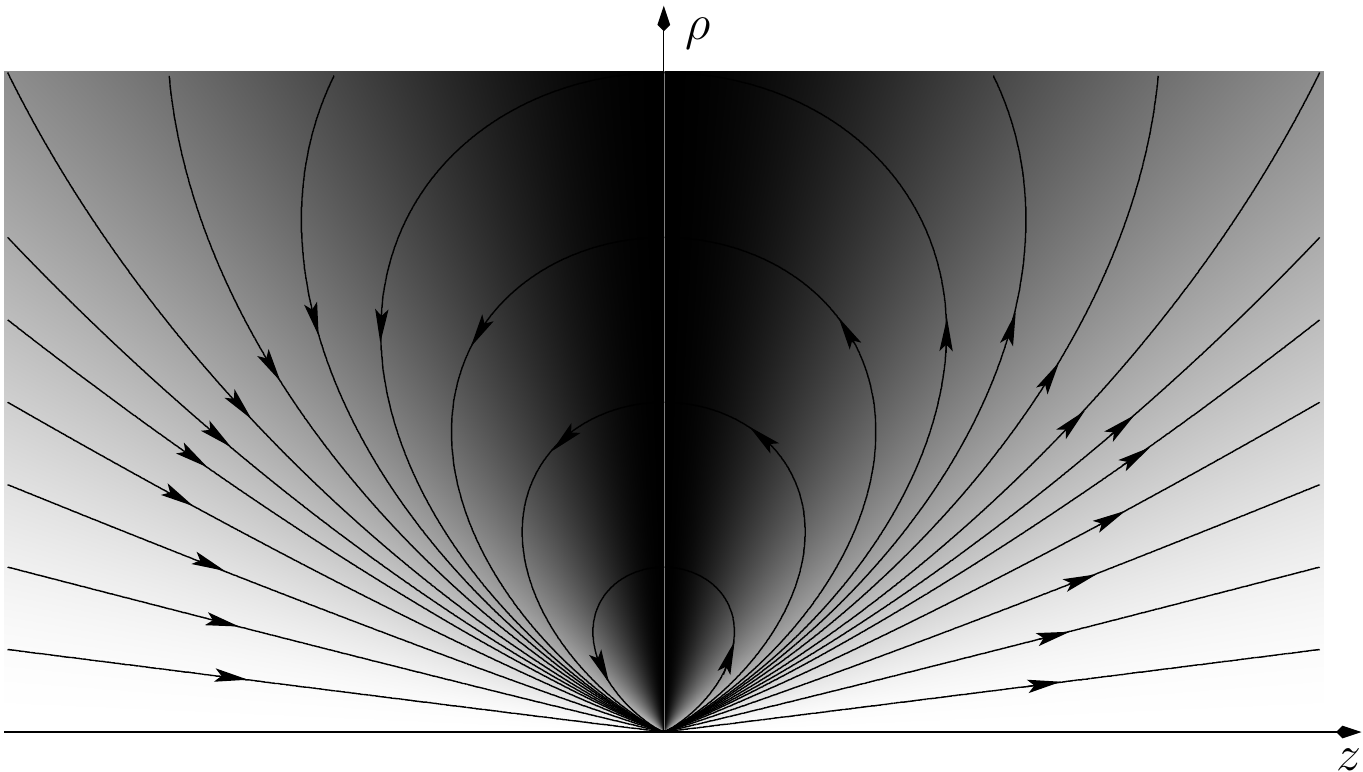}
\caption{Phase velocity field of the paraxial Green's function. Line fields are 
ray trajectories and the level of gray indicates the norm of the phase velocity, being black null phase velocity 
and white wave propagation speed in the medium. The general form of phase velocity confirms that, apparently 
there is not a source at the origin. The wave travels from the $z=0^-$ plane where phase velocity vanishes, 
concentrates in the origin as if there were a source-sink combination and, then travels to the $z=0^+$ 
plane}\label{fig:spfild}
\end{center}
\end{figure}

To finish this example, let us study the field of phase velocity of the paraxial Green's function. The objective is 
to see if the phase of the solution has relevant information. From the previous example, we found that phase velocities 
below the propagation speed in the medium were natural in this scheme of propagation. The phase velocity is a field that 
can be defined as
\begin{equation}\label{eqn:defVf}
\mathbf{v}_p=\omega\frac{\nabla\angle\{\tilde\psi\}}{\left\|\nabla\angle\{\tilde\psi\}\right\|^2}.
\end{equation}
In this case,
\begin{equation}
\angle\{\tilde\psi\}=k\left(z+\frac{\rho^2}{2z}\right);
\end{equation}
and therefore, we obtain
\begin{equation}
\mathbf{v}_p=c\left(1+\frac{\rho^4}
{4z^4}\right)^{-1}\left[\frac{\rho}
{z}\mathbf{e}_\rho+\left(1-\frac{\rho^2}{2z^2}\right)\mathbf{e}_z\right].
\end{equation}
By comparing this last expression to Eq. (\ref{eqn:avmdg}), it can be seen that under high frequency conditions 
$\mathbf{v}_p$ is antiparallel to $\langle\mathbf{p}\rangle$ as was predicted by Eq. (\ref{eqn:cond1}). This feature can 
also be seen in Fig.\ref{fig:spfild}.

The comparison of the field lines in Figs. \ref{fig:AMPGreen} and \ref{fig:spfild} not only shows that, under high 
frequency condition, momentum density and phase velocity are antiparallel, but also where weak dependence of the 
amplitude function $A(\mathbf{r})$ requirement holds in this example. Note that phase velocity in the paraxial Green's 
function also fulfills
\begin{equation}
\|\mathbf{v}_p\|\leq c,
\end{equation}
as happened with paraxial plane waves. 

\subsection{Fundamental Gaussian beam}
\begin{figure}[!htbp]
\begin{center}
\includegraphics[width=8.6cm]{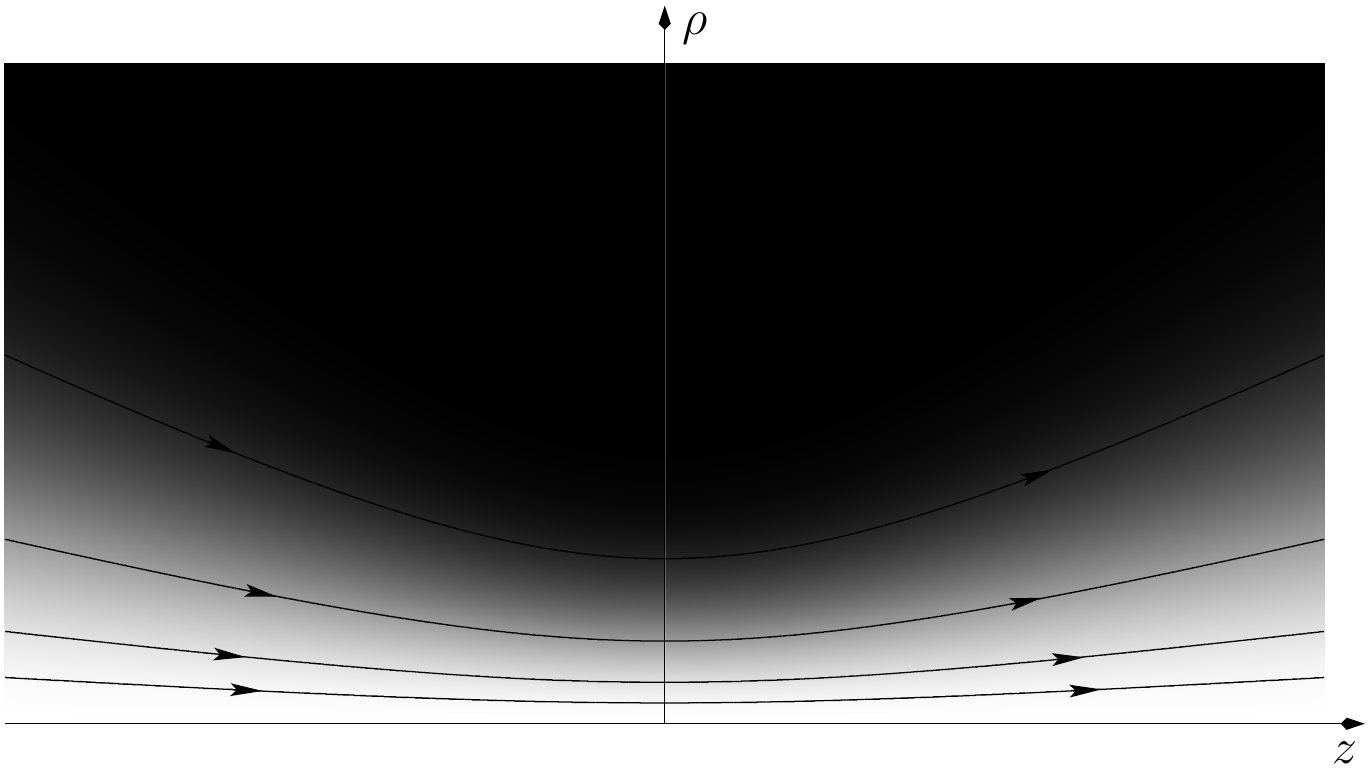}
\caption{Example of energy density flow of a Gaussian beam. The level of gray indicates the value of 
$\langle\mathbf{S}\rangle$: the lighter gray, the higher values. It is straightforward to see from Eq. (\ref{eqn:AvS}) that 
field lines in the $\rho-z$ plane are one branch hyperbolas.}\label{fig:ASrG}
\end{center}
\end{figure}

\begin{figure}[!htbp]
\begin{center}
\includegraphics[width=8.6cm]{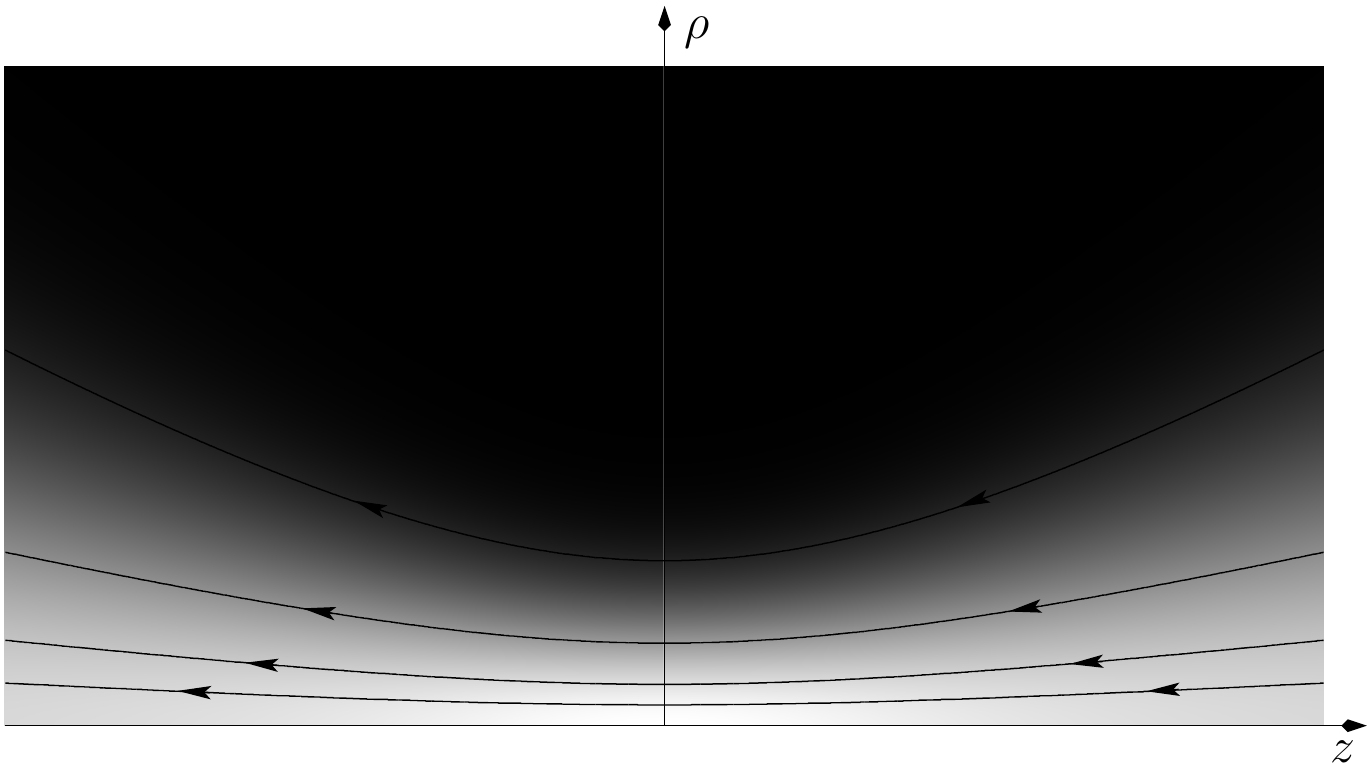}
\caption{Example of momentum density of a Gaussian beam. The level of gray indicates the value of 
$\langle\mathbf{p}\rangle$: the lighter gray, the higher values.}\label{fig:AMrG}
\end{center}
\end{figure}

The phasor associated to the fundamental paraxial Gaussian beam can be written in terms of the Rayleigh distance 
$b$ as follows
\begin{equation}\label{eqn:pGB}
\tilde\psi=\frac{1}{z-ib}\exp\left[ik\left(z+\frac{1}{2}\frac{\rho^2}{z-ib}\right)\right].
\end{equation}
%The generalized velocity can be written as
%\begin{equation}
%\mathbf{V}=ik\,\tilde\psi\left(\frac{x}{z-ib}\mathbf{e}_x+\frac{y}{z-ib}\mathbf{e}_y+\mathbf{e}_z\right).
%\end{equation}
Thus, the complex energy density flow is found by applying Eq. (\ref{eqn:defSc}) as
\begin{equation}\label{eqn:ScG}
\mathbf{S}_c=\frac{\omega}{2}k|\tilde\psi|^2\left(\frac{x}{z-ib}\mathbf{e}_x+
\frac{y}{z-ib}\mathbf{e}_y+\mathbf{e}_z\right).
\end{equation}
This solution, unlike the previous examples, has reactive energy density flow,
\begin{equation}
\mathbf{Q}=\Im\{\mathbf{S}_c\}=\frac{k^2c}{2}|\tilde\psi|^2\frac{b\rho}{z^2+b^2}\mathbf{e}_\rho.
\end{equation}
From Eq. (\ref{eqn:ScG}), the average energy density flow is
\begin{equation}\label{eqn:AvS}
\langle \mathbf{S}\rangle=\Re\{\mathbf{S}_c\}=\frac{k^2c}{2}|\tilde\psi|^2
\left(\frac{\rho z}{z^2+b^2}\mathbf{e}_\rho+\mathbf{e}_z\right).
\end{equation}
A sketch of the average energy density flow can be seen in Fig. \ref{fig:ASrG}
The average momentum density can be calculated as well, yielding
\begin{eqnarray}
\langle\mathbf{p}\rangle=-\frac{k^2}{2c}&|\tilde\psi|^2&\left\{\frac{\rho z}{z^2+b^2}\left(1+\frac{1}{2}
\frac{\rho^2}{z^2+b^2}\right)\mathbf{e}_\rho+\ldots\phantom{\left(\frac{1}{1}\right)^2}\right.\nonumber\\
\ldots&+&\left.\left[1+\frac{2kb-1}{k^2}\frac{1}{z^2+b^2}-
\left(\frac{1}{2}\frac{\rho^2}{z^2+b^2}\right)^2\right]\mathbf{e}_z\right\},
\end{eqnarray}
which is a rather cumbersome expression. An example of momentum density of this solution can be seen in Fig. 
\ref{fig:AMrG}. The resemblance with energy density flow is rather high in contrast to what happened in the previous 
example.

\begin{figure}[!htbp]
\begin{center}
\includegraphics[width=8.6cm]{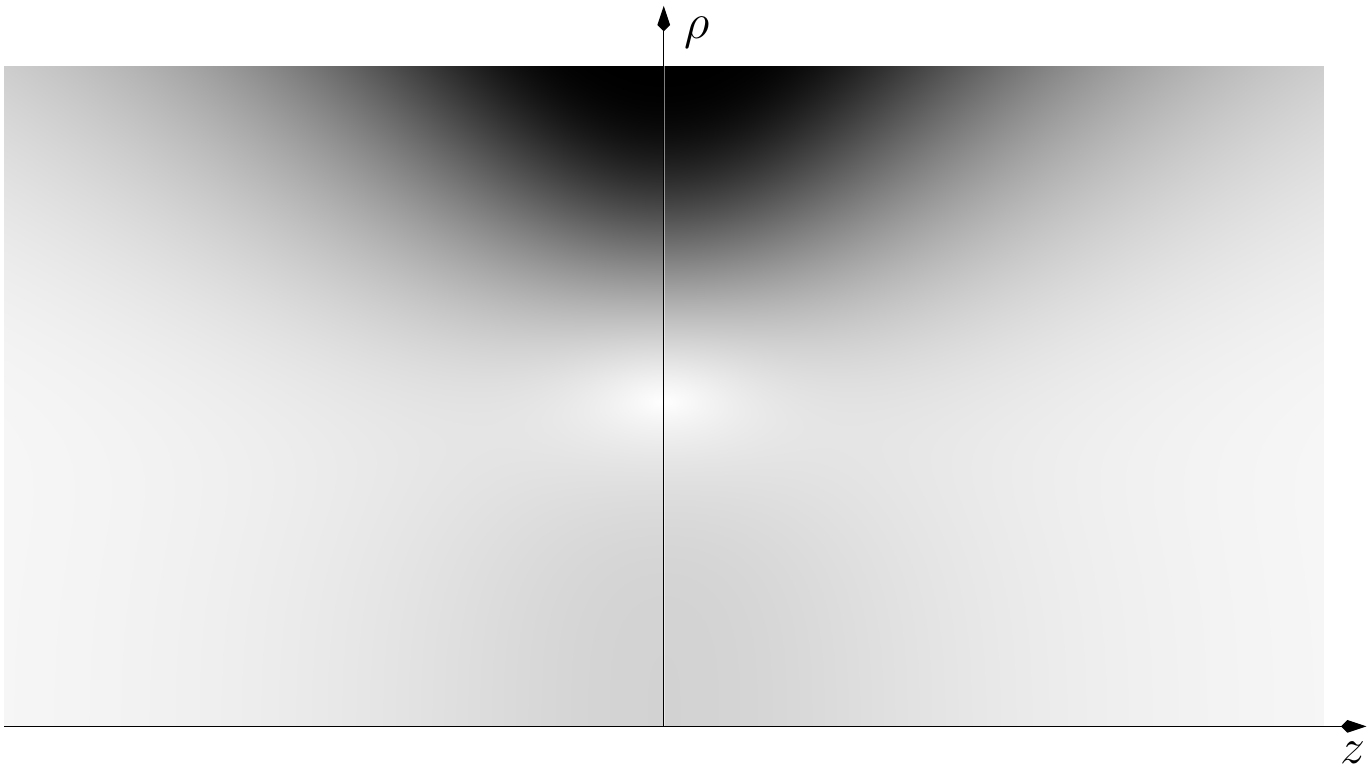}
\caption{Quality in an example of Gaussian beam. The level of gray indicates the quality being the lighter gray, the better 
quality.}\label{fig:DrGB}
\end{center}
\end{figure}

In any case, as in the previous example, we can extract the average energy 
density flow dependence in the momentum density, yielding the vector $\mathbf{D}$ that measures the 
quality of the solution at each point,
\begin{equation}
\mathbf{D}=\frac{k^2c}{2}|\tilde\psi|^2 
\left\{\frac{-\rho^3z}{2(z^2+b^2)^2}\mathbf{e}_\rho+\left[\frac{1-2kb}{k^2}\frac{1}{z^2+b^2}+
\left(\frac{1}{2}\frac{\rho^2}{z^2+b^2}\right)^2\right]\mathbf{e}_z\right\}.
\end{equation} 
Inasmuch as $\mathbf{D}$ can be neglected with respect to $\langle\mathbf{S}\rangle$, the Gaussian beam can be 
considered, in principle, as a fair approximation to the full wave propagation model. At first glance, $\mathbf{D}$ vanishes 
if $z\gg\rho$ and $k\gg 1$, which are the classical ``near the axis'' and ``high frequency'' approximations, respectively. 
Nonetheless, this solution is  required to represent a beam at its waist: in the $z=0$ plane, under 
high frequency conditions, $\mathbf{D}$ can be neglected with regard to $\langle\mathbf{S}\rangle$ for some 
``low'' values of $\rho$ but not the whole plane. All these issues are shown in Fig. \ref{fig:DrGB} where the parameter in 
Eq. (\ref{eqn:defQ}) is taken as a quality measure for each point.
%$\frac{\|\mathbf{D}\|}{\|\langle\mathbf{S}\rangle\|}$   
It is noteworthy that points near the origin in Fig. \ref{fig:DrGB} are not within the lighter ones in this example where the 
high frequency condition is not fulfilled. If the high frequency condition is met, all the points ``near'' $\rho=0$ in 
the $z=0$ plane will raise their quality. Clearly, only a centered disk at the origin in the $z=0$ plane can be considered as a 
good approximation for propagation purposes but not the entire plane.

\begin{figure}[!htbp]
\begin{center}
\includegraphics[width=8.6cm]{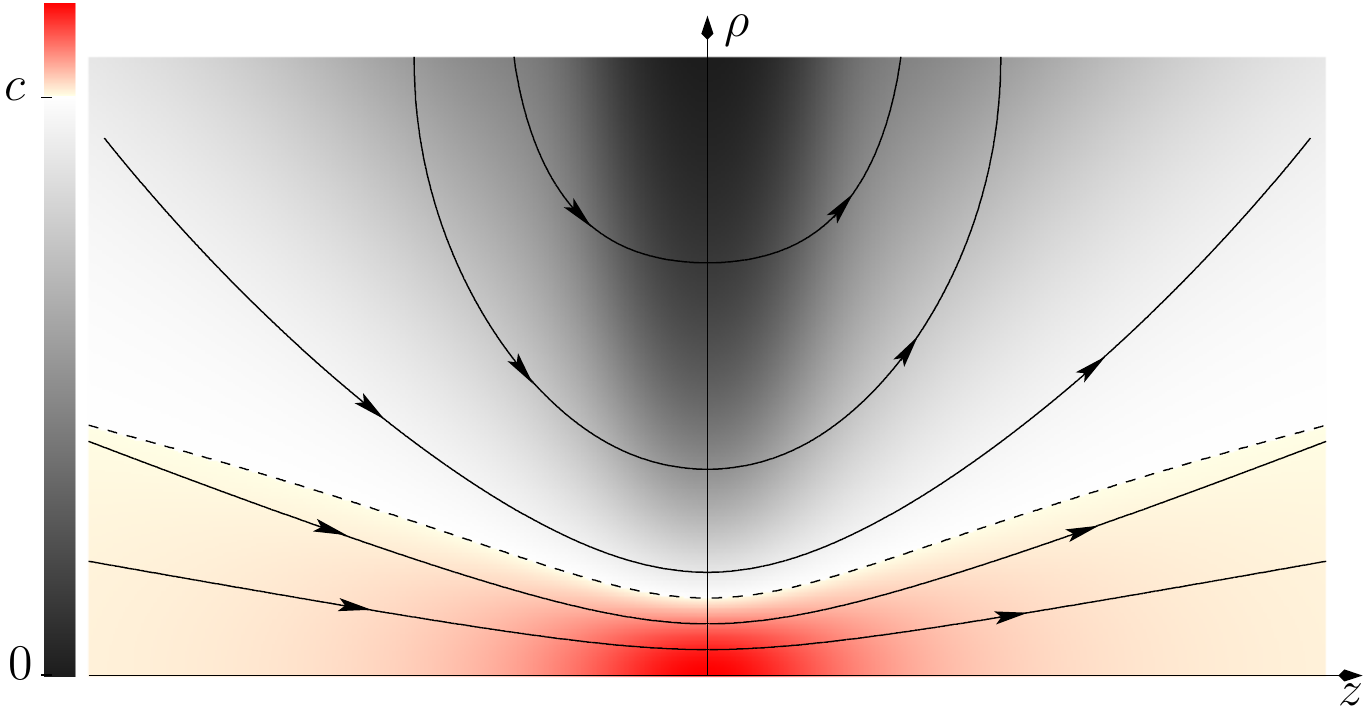}
\caption{(Color) Example of phase velocity of a Gaussian beam. Phase velocity below $c$ is shown in gray as in the 
previous example, while values over $c$ are drawn in red. The locus where $\|\mathbf{v}_p\|=c$ is shown in a 
dashed line.}\label{fig:vfgb}
\end{center}
\end{figure}

The study of the phase velocity is worth it in this example also. By taking the definition in Eq. (\ref{eqn:defVf}), the 
gradient of the phase of the Gaussian beam in Eq. (\ref{eqn:pGB}) is needed to obtain the field of phase velocity. Thus, in 
this example,
\begin{equation}
\nabla\angle\{\tilde\psi\}=k\left\{\frac{z\rho}{z^2+b^2}\mathbf{e}_\rho
+\left[1+\left(\frac{\rho^2}{2}-\frac{b}{k}\right)\frac{1}{z^2+b^2}
-\frac{z^2\rho^2}{(z^2+b^2)^2}\right]\mathbf{e}_z\right\},
\end{equation}
which yields
\begin{equation}
\mathbf{v}_p=c\frac{\frac{z\rho}{z^2+b^2}\mathbf{e}_\rho
+\left[1+\left(\frac{\rho^2}{2}-\frac{b}{k}\right)\frac{1}{z^2+b^2}
-\frac{z^2\rho^2}{(z^2+b^2)^2}\right]\mathbf{e}_z}{\frac{z^2\rho^2}{(z^2+\rho^2)^2}+\left[1+\left(\frac{\rho^2}
{2}-\frac{b}{k}\right)\frac{1}{z^2+b^2}
-\frac{z^2\rho^2}{(z^2+b^2)^2}\right]^2}.
\end{equation}
The evaluation of the phase velocity at the axis is more straightforward,
\begin{equation}
\left.\mathbf{v}_p\right|_{\rho=0}=c\frac{z^2+b^2}{z^2+b^2-\frac{b}{k}}\mathbf{e}_z.
\end{equation}
From this last equation, it is easy to see that
\begin{equation}
\left.\|\mathbf{v}_p\|\right|_{\rho=0}>c.
\end{equation}
In other words, near the axis, the phase velocity only tends to the propagation speed in the medium 
asymptotically. Let us evaluate phase velocity at the plane of the beam waist ($z=0$), obtaining
\begin{equation}
\left.\mathbf{v}_p\right|_{z=0}=\frac{c}{1+\frac{\rho^2}{2b^2}-\frac{1}{kb}}\mathbf{e}_z.
\end{equation}
From this last equation, one may see that 
\begin{equation}
\left.\|\mathbf{v}_p\|\right|_{\rho=w_0,z=0}=c,
\end{equation} 
where $w_0=\sqrt{\frac{2b}{k}}$ is the parameter that defines the beam width at its waist. Phase velocity at points of the 
plane out of this circumference is lower than $c$ whilst, at points contained by the circumference, the phase velocity will be 
higher than the propagation speed in the medium. An example of the field of phase velocity is shown in Fig. \ref{fig:vfgb}.
Note that the locus $\|\mathbf{v}_p\|=c$ does not collapse into the $z$ axis. % Since:
%\begin{equation}
%\rho=\frac{\left[\left(4k^2b^4+4k^2b^2-16kb^3+k^2+8kb+16b^2\right)^{\frac{1}{2}}
%+2b^2k+k+4b\right]^{\frac{1}{2}}}{2bk^{\frac{1}{2}}}z+\mathcal{O}\left(z^{-1}\right)
%\end{equation}
Phase velocity above the propagation speed in the medium is a feature that did not arise in the previous examples. 
Nonetheless, as it happens in the full wave propagation framework, the presence of reactive energy is associated to phase 
velocity greater than the propagation speed in the medium. Note that, on account of the properties of the paraxial Green's 
function with respect to the sources, unlike Helmholtz waves, reactive energy is not associated to a given source 
distribution, either real or virtual. It is possible to reproduce the behavior of a Gaussian beam around its axis in a solution 
of the Helmholtz equation, giving rise to a complex source point solution \cite{CouBel'81, GMGD'10}. This 
reconstruction method is a valid one with the criteria exposed in \cite{Mahillo'16}. But, the complex source 
point solution has to be regularized near the ring determined by the branch point of the complex distance function 
\cite{Kaiser'03,TK'15}. This fact shows that the reactive power in the paraxial framework does not present a problem in 
terms of regularity of the solutions but its counterpart in the Helmholtz scheme presents some difficulties. 

\section{Conclusions}

The application of the \emph{slowly varying envelope approximation} to the d'Alembert equation results in another 
hyperbolic problem that still holds invariant quantities in terms of energy and momentum. It has been shown that these  
quantities do not follow the expressions of their counterparts in the full wave propagation scheme. The comparison of these 
invariants with the expected behavior in terms of the full propagation framework allows us to obtain quality parameters of 
paraxial solutions. We have built a vector one that takes into account the momentum density and energy density flow. 
Although the quality is measured locally, a global parameter could be obtained by integration. Nonetheless, our local 
measure of quality shows that the high-frequency condition turns out to be relevant. We have gained new insights with 
regard to the behavior of paraxial solutions: rays (defined by phase fronts) are parallel neither to energy flow nor to
momentum. Phase velocity can be either higher or lower than the propagation speed in the medium. Phase 
velocity above the propagation speed in the medium is associated with the presence of reactive energy, as happens in the 
d'Alembert equation, while phase velocity below the propagation speed seems to be ``natural'' in paraxial 
propagation. Therefore, general phase matching when reconstructing a solution form a paraxial one is a delicate procedure. 
The study of paraxial Green's function in time harmonic regime showed that any problem with sources turns 
into a simple propagation problem. Altogether, the paraxial propagation framework proposes a sourceless framework where 
the medium seems to be more dispersive than in the full propagation scheme. The study of the Gaussian beam shows that, 
as a solution it may have a remarkable high quality region around its axis. The analysis of the phase velocity shows that 
it exceeds the propagation speed in the medium in the high quality region. Although the particular feature with regard to 
the sources in paraxial solutions does not present a problem, it follows that a solution of the d'Alembert 
equation may need a non-physical distribution of sources in order to maintain that phase speed far from them. 

\begin{acknowledgments}
This work is supported by the Ministerio de Economía y Competitividad
(MINECO) (MINECO/FEDER, EU) (TEC2015-69665-R). The authors want to thank Prof. Carlos Dehesa-Mart\'{\i}nez not 
only for his help in this article but also for his knowledge,  friendship and patience during the last years. Without his 
guidance, the good part of this paper would have never been possible.
\end{acknowledgments}

\appendix
\section{Summary of results for the d'Alembert equation}
A Lagrangian density associated to the d'Alembert equation
\begin{equation}\label{eqn:Daleq}
\nabla^2\psi-\frac{1}{c^2}\partial_{tt}\psi=0
\end{equation}
can be found,
\begin{equation}
\mathscr{L}=\frac{1}{2}\frac{(\partial_t\psi)^2}{c^2}
-\frac{1}{2}\left[(\partial_x\psi)^2+(\partial_y\psi)^2+(\partial_z\psi)^2\right].
\end{equation}
Thus, %the energy density is,
%\begin{equation}\label{eqn:UFW}
%U=\frac{1}{2}\left\{\frac{(\partial_t\psi)^2}{c^2}
%+\left[(\partial_x\psi)^2+(\partial_y\psi)^2+(\partial_z\psi)^2\right]\right\};
%\end{equation}
%and 
the energy density flow is
\begin{equation}\label{eqn:SFW}
\mathbf{S}=-\partial_t\psi\,\nabla\psi.
\end{equation}
The phasor associated to a time harmonic solution of the d'Alembert equation must fulfill the well-known Helmholtz 
equation
\begin{equation}
\nabla^2\tilde\psi+k^2\tilde\psi=0.
\end{equation}
In this case,% the energy density is 
%\begin{equation}\label{eqn:UHelm}
%U=\frac{1}{4}\left[k^2|\tilde\psi|^2+\left\|\nabla\tilde\psi\right\|
%^2+\Re\left\{\left(\nabla\tilde\psi\cdot\nabla\tilde\psi^* 
%-k^2\tilde\psi^2\right)e^{-i2\omega t}\right\}\right]
%\end{equation}
%Hence, the average energy density in time harmonic solutions depends only on the phasor
%\begin{equation}
%\langle U\rangle=\frac{1}{4}\left(k^2|\tilde\psi|^2+\left\| \nabla\tilde\psi\right \|^2\right).
%\end{equation}
the energy density flow depends only on the phasors as
\begin{equation}
\mathbf{S}=\frac{\omega}{2}\left(\Im\left\{\tilde\psi^*\nabla\tilde\psi\right\}-\Im\left\{\tilde\psi\nabla\tilde\psi\,
e^{-i2\omega t}\right\}\right).
\end{equation}
The average energy density flow is
\begin{equation}\label{eqn:Savarm}
\langle\mathbf{S}\rangle=\frac{\omega}{2}\Im\left\{\tilde\psi^*\nabla\tilde\psi\right\}.
\end{equation}
This leads to the definition of the complex energy density flow as
\begin{equation}
\mathbf{S}_c=-i\frac{\omega}{2}\tilde\psi^*\nabla\tilde\psi,
\end{equation}
%where
%\begin{equation}
%P=i\omega\tilde\psi
%\end{equation}
%is the generalized pressure phasor and
%\begin{equation}
%\mathbf{V}=\nabla\tilde\psi
%\end{equation}
%is the generalized velocity phasor.
It follows from its construction that the real part of $\mathbf{S}_c$ is the average energy density flow. The imaginary part 
gives the reactive energy density associated to the solution whose phasor is $\tilde\psi$.
The momentum density associated to a solution of the d'Alembert equation is obtained as follows from the Noether's 
theorem
\begin{equation}\label{eqn:pfw}
\mathbf{p}=\frac{1}{c^2}\partial_t\psi\,\nabla\psi,
\end{equation}
which is almost equal to the energy density flow in terms of the dependence with $\psi$, see Eq. (\ref{eqn:SFW}). From 
Eqs. (\ref{eqn:SFW}) and (\ref{eqn:pfw}) it is clear that $\mathbf{S}$ and $\mathbf{p}$ are antiparallel vectors.
%The angular momentum density is, by definition,
%\begin{equation}\label{eqn:lfw}
%\mathbf{l}=\mathbf{r}\times\mathbf{p};
%\end{equation}
%what can be also expressed in terms of the energy density flow as
%\begin{equation}\label{eqn:lfwS}
%\mathbf{l}=\frac{1}{c^2}\mathbf{S}\times\mathbf{r}.
%\end{equation}
%Note that the definition of the angular momentum density is Eq. (\ref{eqn:lfw}) and not Eq. (\ref{eqn:lfwS}) although 
%the relation in this last equation is much more popular in electromagnetics.
Momentum can be also calculated explicitly for time harmonic solutions. The procedure is quite similar to $\mathbf{S}$ 
and produces
\begin{equation}
\mathbf{p}=\frac{1}{c^2}
\frac{\omega}{2}\left(\Im\left\{\tilde\psi\nabla\tilde\psi^*\right\}+\Im\left\{\tilde\psi\nabla\psi\,
e^{-i2\omega t}\right\}\right).
\end{equation}
From this last equation, the average momentum is
\begin{equation}\label{eqn:pavaram}
\langle\mathbf{p}\rangle=\frac{1}{c^2}\frac{\omega}{2}\Im\left\{\tilde\psi\nabla\tilde\psi^*\right\}.
\end{equation}

\subsection{The connection of energy density flow and momentum with raytracing techniques}\label{susec:con}
For time harmonic waves, the study of phasors gives an idea of momentum and energy flow. A phasor has the general form 
of
\begin{equation}
\tilde\psi(\mathbf{r})=A(\mathbf{r})\exp[i\Phi(\mathbf{r})],
\end{equation}
where $A(\mathbf{r})$ and $\Phi(\mathbf{r})$ are the amplitude and phase functions respectively. By substituting this last 
expression into Eqs. (\ref{eqn:Savarm}) and (\ref{eqn:pavaram}), it follows that
\begin{equation}
\left.\begin{array}{c}
\langle\mathbf{S}\rangle\\
\langle\mathbf{p}\rangle
\end{array}\right\}\propto|A(\mathbf{r})|^2\,\nabla\Phi(\mathbf{r}).
\end{equation}
Thus,
\begin{equation}
\left.\begin{array}{c}
\langle\mathbf{S}\rangle\\
\langle\mathbf{p}\rangle
\end{array}\right\}\parallel\nabla\Phi(\mathbf{r}).
\end{equation}
This last equation explains why the eikonal equation and raytracing techniques are so important. Relevant information of 
the vector structure of the energy flow and momentum is contained in the phase of the phasor without even assuming a 
high frequency condition. If the application of these techniques gives an accurate description of $\Phi(\mathbf{r})$, by 
applying conservation procedures both momentum and energy flow can be obtained and, by extension, the solution $\psi$.

\end{document}